# Identification of large-scale cellular structures on the Sun based on the SDO and PSPT data

V. I. Efremov, L. D. Parfinenko, and A. A. Solov'ev
Central (Pulkovo) Astronomical Observatory, Russian Academy of Sciences, St. Petersburg, Russia
E-mail: parfinenko@mail.ru

Abstract

Three independent sets of data: i). series of filtergrams obtained in line CaII K (393.416 nm) with the ground-based telescope *Precision Solar Photometric Telescope (PSPT)* of *Mauna Loa Solar Observatory*; ii). series of filtergrams of *Atmospheric Imaging Assembly (AIA) of the Solar Dynamics Observatory (SDO)* in $\lambda$160 nm and iii). series of magnetograms of *Helioseismic and Magnetic Imager (HMI)* of *SDO* have been processed to reveal reliably the existence of spatial cellular structures on the solar photosphere at scale about of 300 arcsec. This scale is intermediate between supergranules and giant cells (~30,000 and ~300,000 kilometers across, respectively). To identify the different spatial structures the tens of two-dimensional power spectra (*2DFFT*) have been averaged. For one-dimensional photometric cross sections of frames, the Fourier power spectra (*FFT*) and wavelet transforms (Morlet 5-th order) have been calculated.

## INTRODUCTION

Model of the solar convection proposed by Simon and Weiss (1968) explained the formation of the photospheric granulation and supergranulation, and predicted the existence of larger structures - giant convection cells that are formed at the base of the convection zone. It was expected that the cells extend deeply into the convection zone to a depth of 200,000 km, have diameters of ~ 300,000km and lifetime of ~ 1 month (Miesch 2005; Nordlund, Stein &Asplund 2009). This scale of solar convection can play an important role in the formation of active regions and sunspots (Weber et al. 2012). Expected contrasts of intensity and velocity of plasma flows in the giant cells are very small, so it was a very difficult observational problem to reveal them surely on the background of granulation (~ 1,000 km in diameter) and supergranulation (~ 30,000 km in diameter); evidently, this problem could not be satisfactorily resolved by ground-based observations. For example, the giant cells were not revealed from observations of radial velocities in the Mount Wilson Observatory, with sensitivity ranging from 3 to 12 m/s (LaBonte et al. 1981; Snodgrass & Howard 1984). Using the images of the Sun in the continuum ($\lambda$525.6 nm), the limit on the photometric contrast of giant cells was estimated as 0.016% or 0.23 K (Chianget et al. 1987).

However, some researches have reported the discovery of giant cells or signatures of their existence. Conclusion about the existence of large-scale structure of the surface velocity field that is larger than the size of supergranules was made by Howard and Yoshimura (1979). The large-scale structure of the distribution of the magnetic fields in the observations with low angular resolution was reported by Bumba (1970). He came to the conclusion that there must be a characteristic size of the order of 150,000 km, corresponding to the giant cells, extending through the entire convection zone. The structures exist during 5-8 revolutions of the Sun and are probably associated with the active longitudes.

Using spectroheliograms in line FeI 543 nm, Restano & Bertelli (1992) found a large-scale motions in the solar convection zone, which they have associated with a 5-minute oscillations. Power spectra calculated for the cuts along and across the meridian give the peaks at scale about of 340" × 270". Beck et al. (1998) found some long-lived cell of velocities, which are perhaps the echo of giant convection cells extending to 40-50 degrees in longitude and less than 10 degrees of latitude. Using the Doppler velocity observations obtained with the instruments of Global Oscillation Network Group (GONG), Hathaway et al. (1996) have measured directly the nearly steady flows in the solar photosphere. A meridional circulation with a poleward flow of about 20 meters per second has been clearly revealed in this study. Several characteristics of the surface flows suggest the presence of large convection cells. This was recently confirmed by the processing of HMI/SDO doplerograms. By tracking the motions of supergranules, it was found evidence for giant cellular flows that persist for months. As expected from the effects of the Sun's rotation, the flows in these cells are clockwise around high pressure in the north and counterclockwise in the south and transport angular momentum toward the equator, maintaining the Sun's rapid equatorial rotation (Hathaway at el. 2013).

Line-of-sight magnetograms from the *Helioseismic and Magnetic Imager* (HMI) of the Solar Dynamics Observatory (SDO) were analyzed by Scott W. McIntosh *et al.* (2014) using a diagnostic known as the "Magnetic Range of Influence," or MRoI. This techniques exhibits four scales: a scale of a few megameters (granulation), a scale of a few tens of megameters that can be associated with super-granulation, a scale of many hundreds to thousands of megameters that can be associated with coronal holes and active regions, and a hitherto unnoticed scale that ranges from 100 to 250 Mm. This scale appeared in MRoI maps as well-defined, spatially distributed, concentrations, and authors named it as "g-nodes." They consider this final scale as an imprint of the (rotationally-driven) giant convective scale on photospheric magnetism.

Probably the energy output on the solar surface and the organization of the energy in the different structures is controlled by discrete scales of solar convection. This space structuring, of course, reflects both surface and deep processes occurring on the Sun. Various schemes have been proposed for the structural hierarchy of the solar convection: Ikhsanov (1970), McIntosh (1992) etc. Table 1 shows one of the conventional schemes of hierarchical structures on the Sun (see. Getling A., http://www.magnetosphere.ru/~avg/index.files/lectures.htm)

Table 1. Hierarchy of cellular structures on the Sun

| Type of cells | Sizes (Mm) | Life time | Horizontal velocities | Who discovered |
|---|---|---|---|---|
| Granules | 0.25-2 | 8-15 min | 1-2 km/s | Herschel (1801 г) |
| Mesogranules | 5-10 | 2-10 hours | 0.4-1 km/s | November *et al.* (1981) |
| Supergranules | 20-30 | >1day | 300-500 m/s | Hart (1956), Leighton *et al.* (1962) |
| Giant sells | hundreds | > 100 days | 4 m/s | Simon & Weiss (1968) Beck *et al.* (1998) Hathaway at el. (2013) |

McIntosh and Wilson (1985) suggested the existence of cells intermediate between supergranules and giant cells, occupying about 5 supergranules.

The aim of this work is to search for and identification of structures much larger than supergranules, which, apparently, can be compared with the intermediate cells of McIntosh and Wilson or with the giant ones.

The first attempts to detect large-scale structures on the Sun we have taken in 1992. Spectral methods were used to analyze the full solar disk filtergrams obtained in the CaII K (Efremov, Parfinenko, 1992). Figure 1 shows a two-dimensional power spectrum calculated by the algorithm of N.M.Branner (see Press et al. 1987). It was obtained by summing the individual power spectra for 40 filtergrams of the Sun in the CaII K line for the elements of area 970 "× 970". The harmonics along the solar equator are indicated on X-axis, and harmonics along the meridian are marked on the vertical Y- axis (Code grayscale, 256 levels, is the amplitude of the corresponding harmonic).

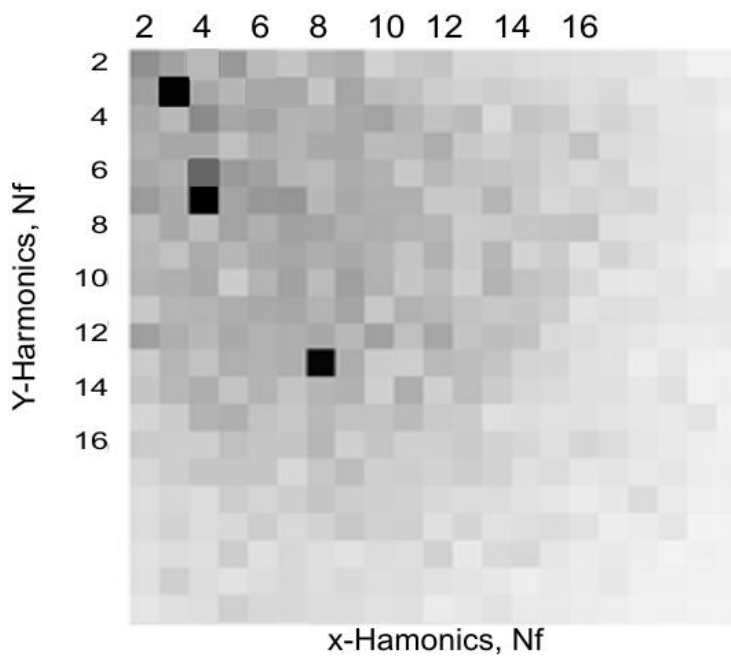

Figure 1: The two-dimensional power spectrum of the brightness field in line CaII K.

In this case, we can identify the following harmonics: in axis Y (along the meridian) the numbers of harmonics with excess of power ($N_f$) are 3, 7, 13; and on the horizontal axis X (along the equator) such numbers are 3, 4, 8. Harmonic number $N_f$ is connected with the spatial scale T" by the following formula: $T" = \frac{L"}{N_f}$, where L" is the length of the scan (in this case, L" = 970").

Thus, in this processing the following dimensional structures: 112"× 69", 223"× 127" and 297" × 297" were clearly manifested. The first corresponds to the scale of supergranulation, and second and third correspond to the scale of a structure close to the intermediate cells of Macintosh and Wilson. Furthermore, large-scale structures on the scales close to the above were detected during processing a series of doplerograms and spectroheliograms by methods of spatial filtering of images (Parfinenko, 1998).

In this paper, we continue to search for a large-scale structures by treatment of various images of the Sun with methods of spectral analysis, using new ground-based and space data. The work is structured as follows: in Section 2 the observational data and processing technique are described, in Section 3 the obtained results are presented and discussed briefly, the final conclusion is given in the last 5-th Section.

## 2. OBSERVATIONAL DATA AND PROCESSING TECHNIQUE

In this paper we use a variety of observational material of high quality. Initially, we have developed our result obtained in the work (Efremov, Parfinenko, 1996), where the calcium filtergrams have been used. Now we have processed FITS-filtergrams for full solar disk in the line CaII K (393.416 nm) obtained at 15 cm refractor telescope Precision Solar Photometric Telescope (PSPT) of Mauna Loa Solar Observatory. Exposure time in seconds, per frame is 0.0260 s. The PSPT produces seeing-limited full-disk digital (2048x2048) images in [CaII K Narrow Band Core (NBC) (393.4nm, FWHM 0.1nm)](), with an unprecedented 0.1% pixel-to-pixel relative photometric precision. (http://lasp.colorado.edu/pspt_access/#data).

We have used also the series of space-based data: the doplerograms and magnetograms obtained with HMI/SDO and filtergrams obtained with UV-visible channel of AIA(SDO) in line $\lambda160$ nm. (http://jsoc.stanford.edu/ajax/data/dataRequest.html). Wavelength band of this UV-channel is associated with transition region and upper photosphere (http://aia.lmsal.com/public/instrument.htm).

Since the objectives of the present study did not include the getting any time-dependences, we did not use the traditional procedure of stabilization of the image of the selected work area (i.e. of the frame). For the image stabilization, the frame would be taken extended along the meridian with the small size along latitude, which is not suitable in the study of the spatial characteristics of the field.

Therefore, the central idea of the processing in the present work consisted in the following:
- working area (frame), which has the shape of a square, should cover the largest possible surface of the Sun, although subject to boundary effects, side length of a square turns to be markedly smaller than the diameter of the Sun (Figure 1);
- within each observation session the material to be treated (filtergrams, magnetograms...) was mixed randomly to get rid of time-dependences, and then the several independent samples have been selected from the sets obtained after the mixing.
- within each such sample, the two-dimensional power spectra (2DFFT) were calculated for the given frame, and then the obtained spectra were summarized within the sample.

Moreover, for the one-dimensional cuts made along the frame as parallel to the equator and to the central meridian, the Fourier power spectra (FFT) were calculated, and the wavelet transforms (Morlet 5-th order) were performed.

## 3. RESULTS AND DISCUSSION

### 3.1. Processing of AIA/SDO ($\lambda160$ nm) filtergrams

From the set of filtergrams obtained with AIA (SDO), we have used the images obtained in line $\lambda160$ nm, in which the large-scale spatial structure of the upper photosphere is most noticeable. Figure 2 shows the spatial form and size of the working area of the image of the Sun. In device AIA(SDO) a matrix of 4096x4096 px is used, therefore the frame of 2048x2048 px covers a large part of the surface of the visible disk of the Sun.

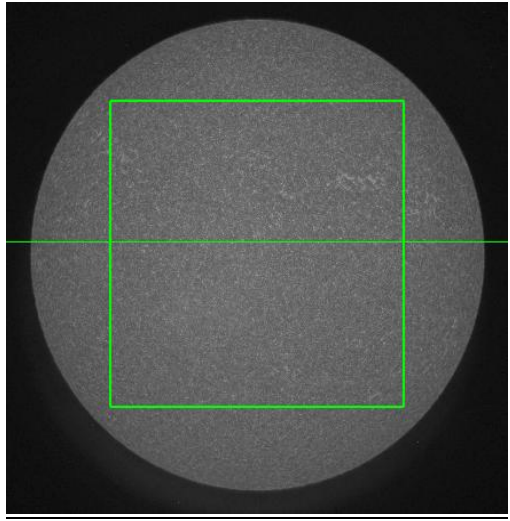

Figure 2. Form of the working area (frame). Size of the SDO matrix is 4096x4096 px, and size of the frame is 2048x2048 px.

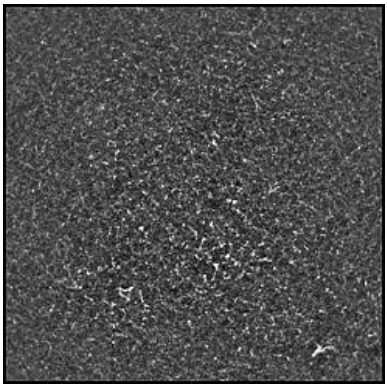
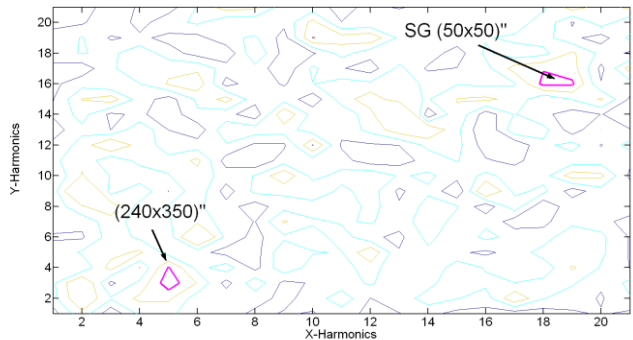

Figure 3a. View of the surface of the Sun in line UV1600A at 2010/09/30 (negative image). Quiet area of size 2048x2048 is shown.

Figure 3b. Contour map of the distribution of the average power of two-dimensional Fourier transform. The spectrum has an excess of power at spatial scales $L_x \times L_y$, where $L_x \sim 240"$, $L_y \sim 350"$, and at scales $(50 \times 50)"$. The map was built by the averaging of 63 2D-spectra.

Figure 3a shows a part of the surface of the Sun derived in line UV1600A, in which the supergranulation cell are clearly manifested. Figure 3b represents the average two-dimensional power spectrum (2DFFT), calculated for the selected area. Averaging was carried out over 63 2D-spectra obtained on frames 2048x2048 px of images of the Sun in line UV1600A. The frame was centred on the center of the solar disk. The excess of power is clearly visible at two spatial scales of [50×50]" and [240×350]". The first is associated with supergranules and the second corresponds to the intermediate or giant cells. Considering 2D contour map, one can notice an important feature: while the structures of the supergranulation scale are almost symmetrical, the larger-scale structures are extended in latitude. The numerical values of the excess harmonics numbers (x: = 5, y: = 3) indicate this fact clearly.

The power spectrum shows the characteristic distances «L» between the structures within this frame, so the real size of the structure «S» may differ from the L depending on the topological filling of the frame field. Information concerning the spatial geometric shape of the studied structure in this case is absent. When the frame is filled completely with the spatial structures, so that their boundaries are in contact, the distance between the centres of the structures are comparable to the size of the structures themselves, i.e. we have: L ~ S. In this case we can talk about the cellular structure of the field, for example, about the cellular convection.

For one-dimensional cuts of the frame made as parallel to the equator and to the central meridian, the power spectra and wavelet transforms were calculated. Here, in Figure 4, the wavelet transform is presented for the one of these cuts (upper panel), and the Fourier spectrum averaged for 50 random cuts is presented in the lower panel . As the wavelet transform and averaged power spectrum, reveal with high confidence an isolated periodic large-scale structure of size ~ 330". This is the same result that was obtained above, using a two-dimensional transformation (2DFFT).

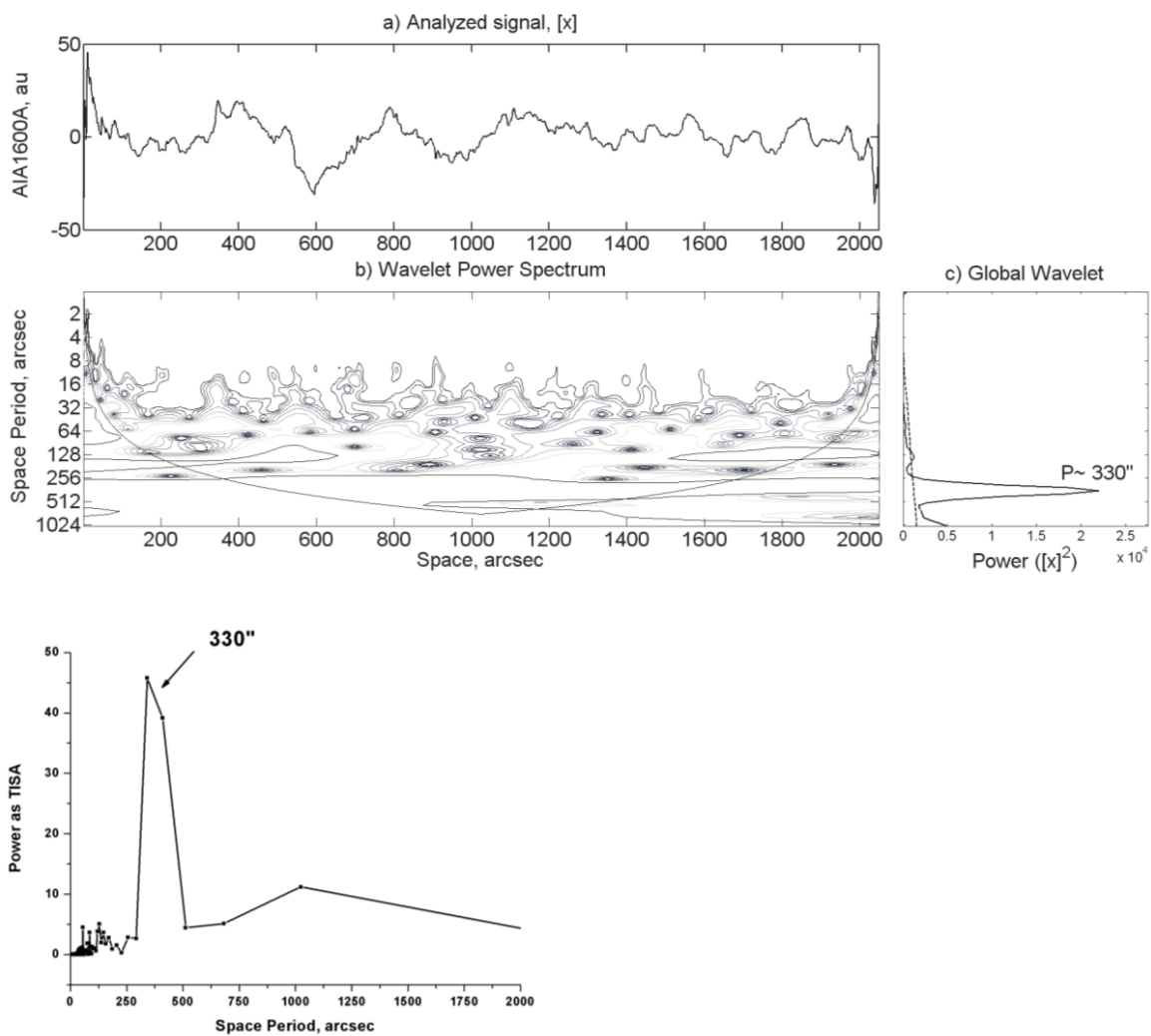

Figure 4. Wavelet transform (Morlet 5-th order) made for the cut along the central meridian (top panel), and the power spectrum (FFT) averaged over 50 cuts (bottom). In both cases, the periodic structure is visible at scale of ~ 330 ".

3.2. Processing of telescope PSPT filtergrams obtained in CaII K (λ160 nm)

A similar study was carried out by us for the filtergrams obtained in the line of CaII K, λ160 nm. Since the size of the matrix of the telescope PSPT is 2048x2048 px, the size of the working area was in this case of 1024x1024 px. Figure 5a shows a surface view of the Sun for this frame in line CaII K, and in Figure 5b the average power spectrum (FFT), obtained for the horizontal and vertical cuts is presented. As can be seen, in this case the excess of power falls not on 5-6 harmonic, as in the previous case, and on 2-3 (order of points counted on the spectrum from left to right), which, of course, reduces the accuracy of the result, but qualitatively it remains the same: the characteristic length scale of the structures remains of the same order and is ~ 250-300 ".

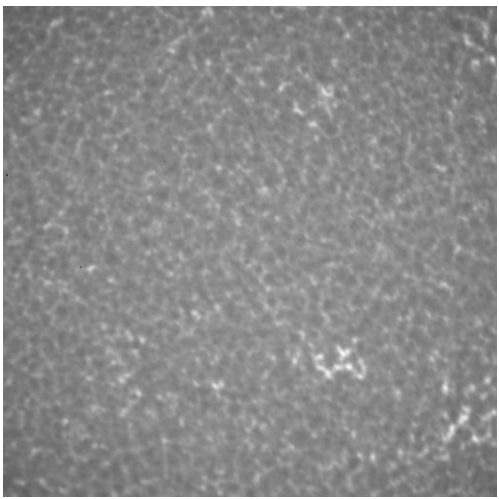
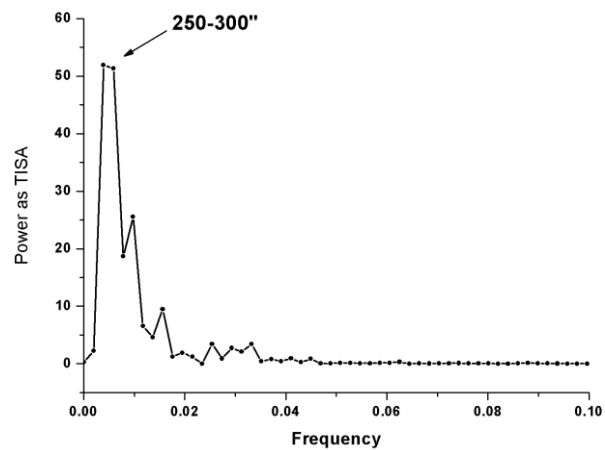

Figure 5a. View of the surface of the Sun in line CaII K at 2005/12/05. The frame size is of 1024x1024 px.

Figure 5b. The average power spectrum (FFT), obtained for the horizontal and vertical cuts.

### 3.3. Processing of HMI/SDO magnetograms

To study the large-scale spatial structure of the "background" magnetic field, we chose the magnetograms of quiet Sun, without exhibiting them a significant activity. To solve this problem, the data of HMI/SDO instrument are well suited. Figure 6 shows the working area (frame) with the size of 2048x2048 px (left) and two-dimensional contour map of the power spectrum (right), respectively. A contour map of the spectrum was obtained by the averaging over hundreds of two-dimensional spectra calculated for the selected magnetograms. Due to a high resolution, as in the magnetic field (~ 10 G) and in space (1 px ~ 0.6 ") we were able to reveal the existence of large-scale structures at scale of giant cells. Despite the noisiness of the spectrum obtained, again the dominant structure of the scale about ~ 300 " stands out clearly.

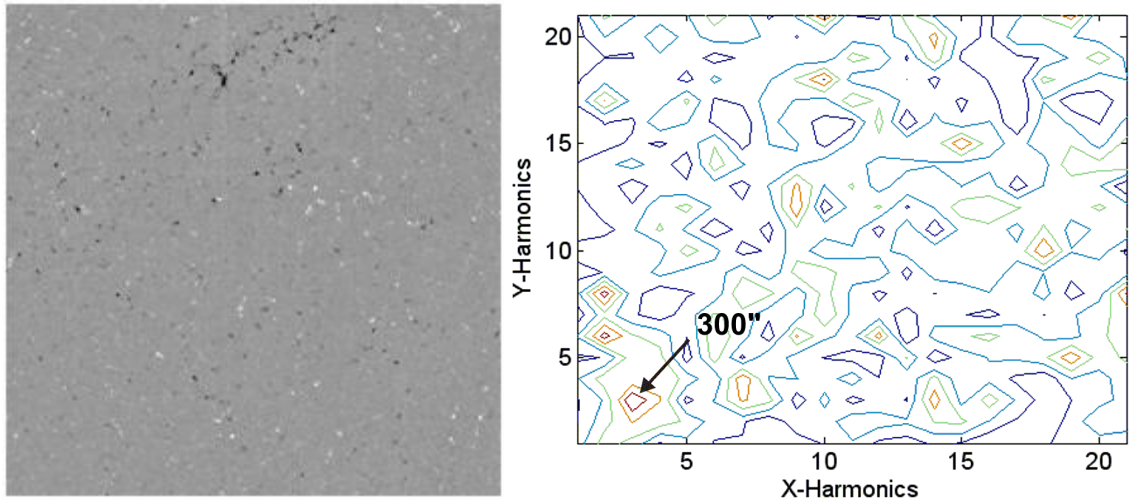

Figure 6. Magnetogram of HMI / SDO, 2048x2048px (2010/10/04) and the resulting two-dimensional contour map of the power spectrum (averaged over hundreds of individual spectra).

4. CONCLUSIONS

For the study of the Sun's images in different spectral ranges, we used a two-dimensional spectral analysis, Fourier analysis and wavelet transform.

The processing of three independent sets of data: filtergrams of ground-based telescope PSPT in CaII K (393.416 nm; space filtergrams of AIA (SDO) in λ160 nm and magnetograms of HMI / SDO has been performed. As a result, the existence in the solar photosphere the spatial cellular structures of scale more than supergranules one (~ 300 ") was revealed with a high degree of confidence.

The similar processing we have made for series of doplerograms obtained with HMI/SDO; despite the high quality of the material, this treatment did not give a reliable results.

This means that the radial velocity in the large-scale structures, we have identified here according to the data above, are very small. The stated accuracy of HMI/SDO doplerograms is better than 13m/s. Therefore, it is possible to take 10 m/s as an upper limit of the radial velocity in the large-scale cells. Since the typical transverse dimension L of the structures we have revealed is about of 300 ", and the characteristic velocity V is admitted to be less than 10 m/s, the time-life of such convective cells, defined as relation L/V, must be greater than 250 days.

These results are generally consistent with existing today conceptions of large-scale convection on the Sun.

ACKNOWLEDGMENTS

We thank the SDO and PSPT teams for the opportunity to use the observation data. This work was supported by the Presidium of the Russian Academy of Sciences, project P-22; the Science School Support Program, project NSh-1625.2012.2, and the Russian Foundation for Basic Research, project 13-02-00714.